\documentclass[prb,10pt,twocolumn,superscriptaddress,aps,amsmath,amsfonts,notitlepage,longbibliography]{revtex4-2}
\usepackage{amsmath,amsfonts,amssymb,dsfont,graphicx,bm}
\usepackage[inkscapelatex=false]{svg}
\usepackage{color}
\usepackage{hyperref}
\usepackage{tikz}
\usepackage{pbox}
\usepackage{balance}

\usepackage{txfonts}

\usepackage{xr}
\usepackage[normalem]{ulem}

\def\be{\begin{equation}}
\def\ee{\end{equation}}
\def\bea{\begin{eqnarray}}
\def\eea{\end{eqnarray}}
\def\bpm{\begin{pmatrix}}
\def\epm{\end{pmatrix}}

\def\diff{\mathrm{d}}

\begin{document}
\title{Proximate quantum spin liquids and Majorana continua in magnetically ordered Kitaev magnets}

\author{Peng Rao}
\affiliation{Physics Department, Technical University of Munich, TUM School of Natural Sciences, 85748 Garching, Germany}
\author{Roderich Moessner}
\affiliation{Max Planck Institute for the Physics of Complex Systems, 01187 Dresden, Germany}
\author{Johannes Knolle}
\affiliation{Physics Department, Technical University of Munich, TUM School of Natural Sciences, 85748 Garching, Germany}
\affiliation{Munich Center for Quantum Science and Technology (MCQST), Schellingstr. 4, 80799 München, Germany}

\date{\today}
\begin{abstract}
We study the spin excitation spectra in magnetically ordered phases proximate to the Kitaev quantum spin liquid (KQSL). Although the low-energy universal features should be governed by the magnetic orders, the \textit{non-universal} high-energy features of the KQSL and adjacent phases can be remarkably similar. Therefore, we study the extended Kitaev model within a Stoner-like theory using Majorana partons, and compute the inelastic neutron scattering (INS) intensities in the random phase approximation. First, we benchmark against the antiferromagnetic (AFM) Heisenberg model and recover the AFM order with linear Goldstone modes. We then explore the phase diagram which agrees qualitatively with previous numerical results. In particular, the Majorana parton theory accurately captures Order-by-Disorder effects in the Kitaev-Heisenberg limit. We also find large INS intensities near the associated high-symmetry Brillouin zone (BZ) points of the magnetic orders. At intermediate and high energies, broad multi-spinon continua emerge across the BZ, providing a distinct mechanism for magnon decay and spectral broadening beyond the conventional multi-magnon decay scenario. Finally, we study the model Hamiltonian of candidate Kitaev material $\alpha$-RuCl$_3$. The zigzag ground state agrees qualitatively with experiments, its stability under external magnetic field also exhibits strong anisotropy in the field directions, and broad scattering continua are recovered similar to those observed experimentally.

\end{abstract}

\maketitle

\section{Introduction}






 Symmetry and conservation laws are ubiquitous and powerful tools to determine the \textit{universal} long-wavelength properties of a physical system. 
On the one hand, the low-energy universal excitations, e.g. Goldstone modes~\cite{halperin1969hydrodynamic}, carry distinct signatures of a given symetry broken phase, and can be experimentally measured by dynamical probes such as inelastic neutron scattering (INS). On the other hand, these probes can also assess intermediate and high energies, where the features are \textit{non-universal} and determined by microscopic correlations of the system. As a result, in systems with multiple competing phases the high energy responses are expected to be similar across the transition points. This is particularly useful in the experimental search for new and exotic quantum states of matter, where one must determine the proximity of a given phase to the target one. 

One prominent example of such exotic phases are quantum spin liquids (QSL), exotic phases of quantum magnets in which conventional magnetic order is destroyed by quantum fluctuations leading to long-range entangled ground states~\cite{savary2016quantum,knolle2019field}. QSL appear in phase diagrams proximate to various magnetically ordered phases, i.e. a prominent example the two-dimensional antiferromagnetic phase in the context of high-temperature superconductivity~\cite{anderson1987resonating,anderson1987resonating-1}. The QSL hosts fractionalized spin excitations as `partons' mapped from spin operators interacting via gauge fields~\cite{lee2006doping}. Thus a given QSL phase corresponds to a `deconfined' phase of partons, whereas in the magnetically ordered phases, single partons are `confined' and cannot be observed as quasiparticles. However, proximate to the QSL phase the confinement length is expected to be long, and partons can still be used to describe higher energy processes as virtual particles. For example, {\it Ho, Muthukumar, Ogata and Anderson} in Ref.~\onlinecite{ho2001nature} used partons to study the broad high-energy spectral of spin excitations in the antiferromagnetic phase of undoped cuprates, based on the 'proximate  QSL' idea. However, in a strongly correlated magnet, there is no general framework to determine which type of partons to use. The choice depends on the type of QSL and needs to be adapted case by case.

However, there is a class of models where the choice of the parton representation is natural, namely for the soluble Kitaev model which hosts neutral Majorana excitations and $\mathbb{Z}_2$ gauge fields~\cite{kitaev2006anyons}, which persist as excitations beyond the soluble point~\cite{knolle2018dynamics}. The extended Kitaev models contain additional spin interactions and are relevant to spin-orbit coupled Mott insulators that are argued to realize a Kitaev quantum spin liquid (KQSL)~\cite{jackeli2009mott}, the most prominent of which being $\alpha$-RuCl$_3$~\cite{plumb2014alpha}. 
Although these materials have magnetic order at low temperatures, they have been argued to be 'proximate' to the KQSL which can be induced by small perturbations to the Hamiltonian~\cite{banerjee2016proximate,banerjee2017neutron,banerjee2018excitations}. For example, in $\alpha$-RuCl$_3$ the low-temperature zigzag phase is destroyed by an in-plane external magnetic field around $7-12$~T at which a KQSL is argued to be realized, before the system becomes a trivially polarized paramagnet. However, despite much research activity, whether the intermediate field phase is a KQSL is not resolved; for reviews see Refs.~\cite{hermanns2018physics,takagi2019concept,motome2020hunting,trebst2022kitaev,rousochatzakis2024beyond,matsuda2025kitaevquantumspinliquids}.

\begin{figure*}[t]
    \centering
    \includegraphics[width=0.95\linewidth]{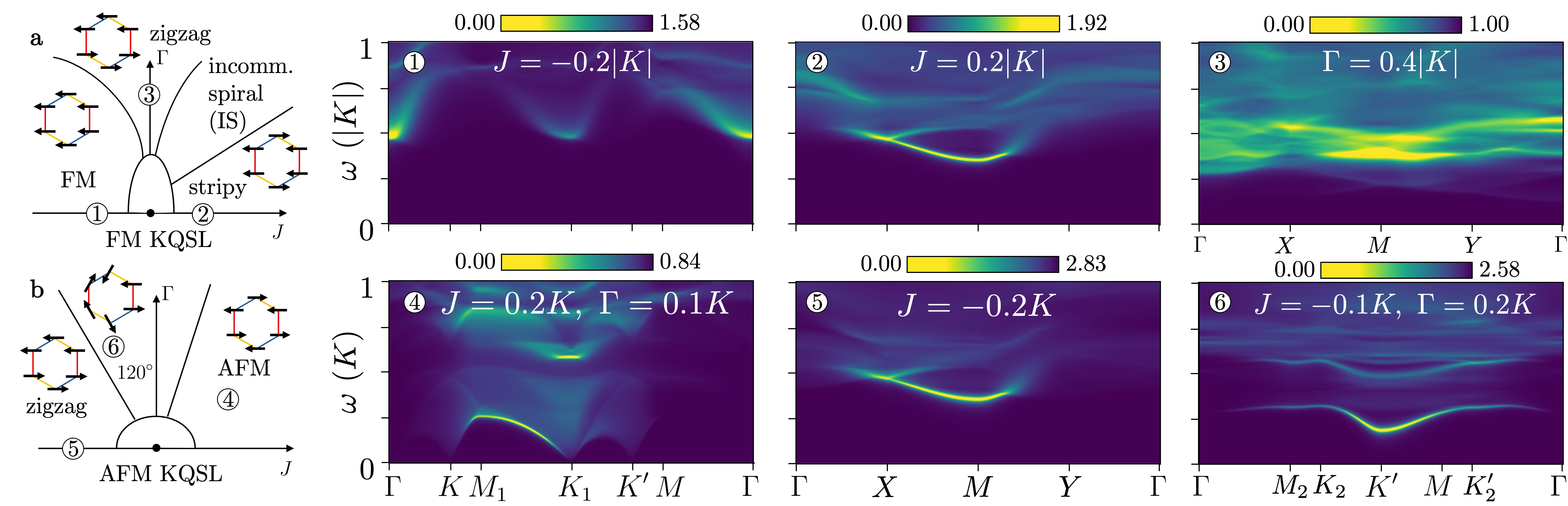}
    \caption{\textbf{Phase diagram and INS intensities of the proximate KQSL in the $\mathbf{KJ\Gamma}$-model.} \textbf{a} (upper panel) near the FM KQSL phase; \textbf{b} (lower panel) near the AFM KQSL phase. The phase diagrams are adapted schematically from Ref.~\cite{rau2014generic}. The momentum path convention is shown in Fig.~\ref{fig:unitcell}. The INS intensity is plotted in logarithmic scale introduced in Sec.~\ref{sec:spin-susceptibility}.}
    \label{fig:phases}
\end{figure*}

The magnetically ordered `proximate KQSL' phase of $\alpha$-RuCl$_3$ also exhibits unusual behaviors. At zero magnetic field, inelastic neutron scattering (INS) experiments found sharp magnon modes coexisting with a large overdamped continuum~\cite{banerjee2017neutron,banerjee2018excitations,sarkis2026intermediate}, also confirmed by DMRG simulations~\cite{winter2017breakdown}. Theoretically, such a continuum can be attributed to the strong frustration of anisotropic spin interactions, which leads to magnon non-conserving decay processes in spin wave theory (SWT)~\cite{zhitomirsky2013colloquium,winter2017breakdown}. An important question is which other framework beyond the SWT scenario could explain magnon damping and decay? In itinerant magnets magnons can be broadened by decaying into particle-hole excitations of fermionic excitations and it is unclear whether a similar scenario can apply in the presence of emergent fermionic spinon excitations. Here, drawing on the foregoing considerations, we elaborate such an alternative scenario in detail: we establish that magnons can be overdamped by the presence of multi-spinon continua of a proximate QSL. It allows us to provide an explanation for the high-energy INS features of a Kitaev magnet by its proximity to the KQSL phase.

Concretely, we study the proximate ordered phases of generic Kitaev magnets in the Majorana representation at zero temperature. In a previous work~\cite{rao2025dynamical}, we employed Majorana mean field theory (MFT) and the random phase approximaion (RPA) to compute the INS intensity of the $KJ\Gamma$-model within the KQSL phases, where the MF magnetization $\mathbf{m}=0$ vanishes at zero magnetic field. In the pure Kitaev limit, we did find a remarkable even quantitative agreement between the RPA theory and the  exact solution for the dynamical spin susceptibility~\cite{knolle2014dynamics,knolle2016dynamics}. Moreover, it was found that non-Kitaev couplings induce sharp paramagnon modes as bound states of partons within the RPA INS intensity. Instabilities towards magnetic orders appear via the condensation of these modes at the corresponding high-symmetry points in the Brillouin zone (BZ). Here, we use the resulting phase diagram as an input and concentrate on the proximate ordered phases solving for non-zero $\mathbf{m}$ on the MF level and then performing self-consistent RPA for the INS intensities. 


\begin{figure*}
    \centering
    \includegraphics[width=\linewidth]{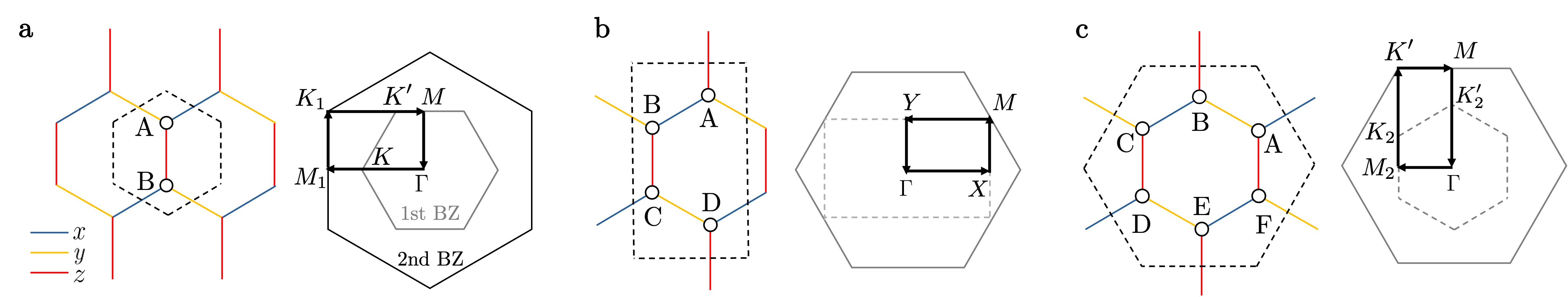}
    \caption{\textbf{Lattice and Brillouin zone (BZ) conventions}. The unit cells are shown in black dashed lines with: \textbf{a} two sublattices (KQSL, FM, AFM); \textbf{b} four sublattices (zigzag, stripy); \textbf{c} six sublattices ($120^\circ$). The first and second BZs of the honeycomb lattice are shown in gray and black solid lines respectively. The BZs of the four- and six-sublattice unit cells are shown in grey dashed lines.}
    \label{fig:unitcell}
\end{figure*}

We first benchmark our method for the AFM Heisenberg model, which is far from any QSL regime on the bipartite honeycomb lattice. Nevertheless, the Majorana RPA reproduces the standard linear dispersion of the gapless Goldstone mode at low energies, with a renormalized spin wave velocity compared to SWT. We then explore the ordered phases of the $KJ\Gamma$-model, where the Majorana MF gives good qualitative agreement with numerical studies on the nature of the phases and the magnetization directions. In the INS intensity, the low-energy magnon modes are located near the high-symmetry BZ points of the corresponding order; they are generally sharp and gapped away from the phase boundary with KQSL. We focus on the Kitaev-Heisenberg limit, where Order-by-Disorder (ObD) effects are prominent~\cite{chaloupka2010kitaev,chaloupka2013zigzag,rau2014generic,sela2014order,chern2017kitaev}. In ObD the classical ground state manifold in linear SWT (LSWT) has a larger accidental degeneracy unrelated to symmetry, resulting in gapless pseudo-Goldstone (pG) modes~\cite{villain1980order,shender1982afm,rau2018pseudo,khatua2023pseudo,rao2025order}. This accidental degeneracy is normally removed by magnon interactions which also gap out the pG modes. Now within our parton-based theory we find that in the Kitaev-Heisenberg model the Majorana theory has better agreement with the numerics on the ground state and the spin gap, which differ drastically from the LSWT results~\cite{chaloupka2010kitaev,chaloupka2013zigzag,rau2014generic,rau2018pseudo}. We attribute the agreement to the fact that Majorana MF does not directly exapnd around the classical spin Ansatz and captures part of the quantum fluctuations. 

Finally, we study the $KJ\Gamma\Gamma'J_3$-model, which is argued to describe the Kitaev candidate material $\alpha$-RuCl$_3$~\cite{moller2025saga}, under external magnetic fields. We emphasize that in our results, both the sharp modes and the continua are generated self-consistently using a single, general framework without additional ad hoc approximations. We argue that our results are in qualitative agreement with recent experiments~\cite{sarkis2026intermediate}.

\section{Results}
Throughout we consider the $KJ\Gamma$-model:
\begin{equation}\label{eq:KJGamma-Hamiltonian}
    H_{KJ\Gamma} =K\sum_{i,j\in \alpha} S^\alpha_iS^\alpha_j+J\sum_{ \text{N.N.}}\mathbf{S}_i.\mathbf{S}_j+   \Gamma \sum_{\substack{i,j \in \alpha \\ \alpha \ne \beta \ne \gamma}} S^\beta_i S^\gamma_j,
\end{equation}
with the Kitaev coupling $K$, the nearest neighbour (N.N.) Heisenberg coupling $J$, and an off-diagonal symmetric coupling $\Gamma$. The honeycomb lattice conventions and BZ details used in the following are given in Fig.~\ref{fig:unitcell}.

\subsection{Phase Diagram}
As established in Ref.~\cite{rao2025dynamical}, within a parton-based RPA theory within the KQSL phase, the latter becomes unstable when the sharp magnon modes condense. Here, we decouple the spin interactions written in the parton basis self-consistently allowing for finite magnetic order. The magnetic orders are shown schematically in Fig.~\ref{fig:phases}, and are discussed in detail in subsequent sections. We briefly note some general features: Majorana MFT gives qualitatively the same magnetic orders as previoulsy found with classical solutions, exact diagonalization and DMRG~\cite{rau2014generic,gohlke2017dynamics} albeit with quantitative differences expected for any MFT. The resulting INS intensities also show distinctive features within the ordered phases. Near the transition points, the magnon modes at the corresponding high-symmetry momenta points are gapped and more broadened compared to the KQSL phase. However as the non-Kitaev couplings further increase, the magnon modes become sharper. In the following, we discuss different limiting cases and points of the rich phase diagram.

\subsection{The Kitaev-Heisenberg model and Order by Disorder}

Let us discuss the results in different parameter regimes in greater detail. We start from the Kitaev-Heisenberg model, i.e. setting $\Gamma=0$ in Eq.~\eqref{eq:KJGamma-Hamiltonian}. Ref.~\cite{rao2025dynamical} considered the pure Kitaev limit using the present approach. Rather surprisingly it was found that the RPA spin susceptibility agreed {\it quantitatively} with the exact solution~\cite{knolle2014dynamics}. Remarkably, we show in the following that also in the opposite `limit' of a pure unfrustrated AFM Heisenberg model, the parton RPA method captures qualitatively the low-energy properties of the ordered system. 

\begin{figure}
    \centering
    \includegraphics[width=\linewidth]{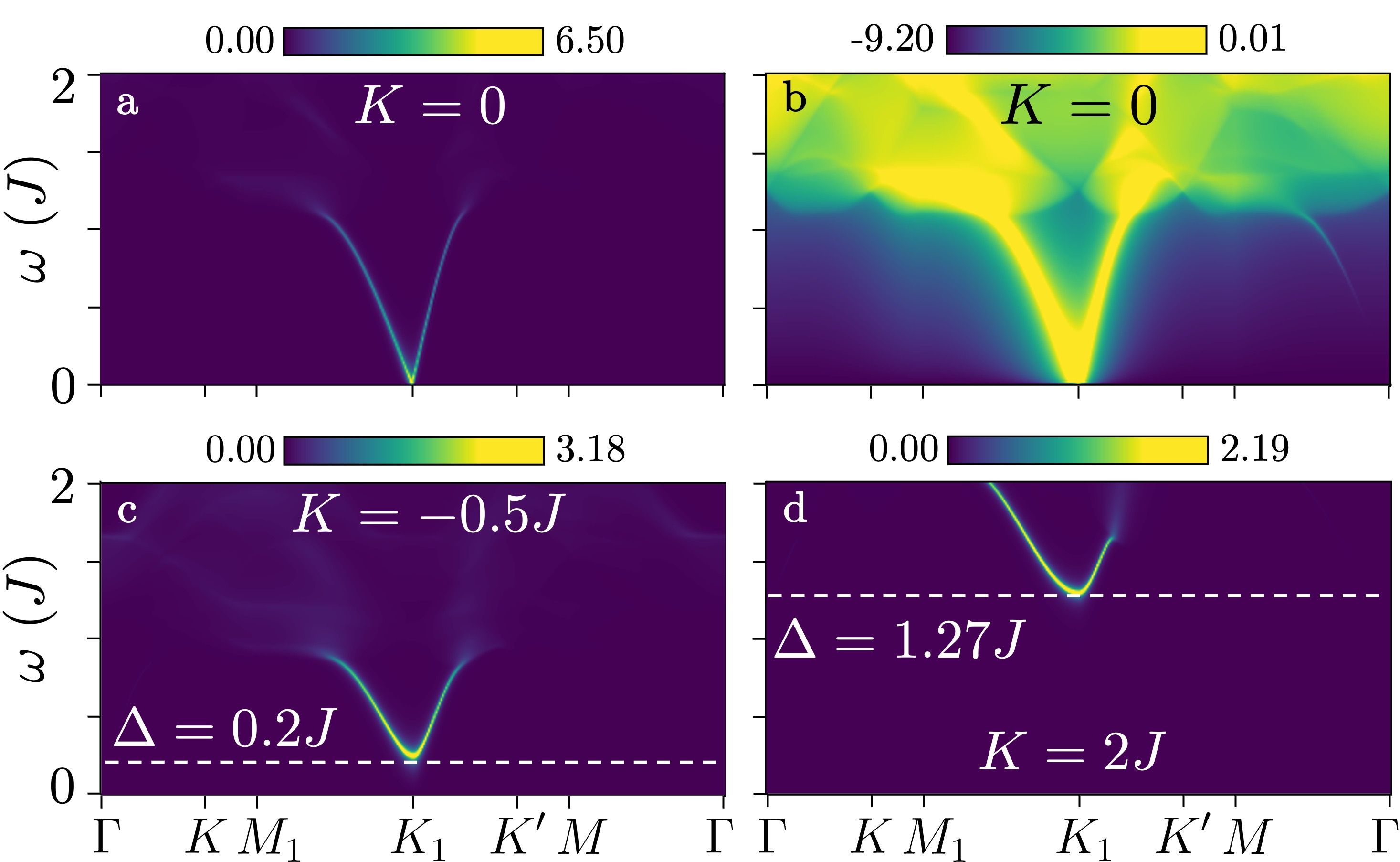}
    \caption{\textbf{INS intensities for the Kitaev-Heisenberg model inside the AFM phase at $J>0$.} The intensities are shown in logarithmic scale introduced in Sec.~\ref{sec:spin-susceptibility} which makes the magnon mode more visible. In \textbf{a}, \textbf{c}-\textbf{d} the off-set value $c=1$ introduced below Eq.~\eqref{eq:INS-intensity}. In \textbf{b} we set $c=10^{-4}$ to highlight the low-intensity features, compared to \textbf{a} which focuses on the sharp gapless Goldstone mode. The apparent anisotropy of the Goldstone mode spin velocity in \textbf{a}-\textbf{b} is due to that the momentum path lengths $M_1\rightarrow K_1$, $K_1\rightarrow K'$ are scaled differently.}
    \label{fig:ObD}
\end{figure}

In particular, we find that the MF magnetic moment directions have full rotational symmetry and the magnon modes are gapless as a result of spontaneous symmetry breaking; see Fig.~\ref{fig:ObD}\textbf{a}-\textbf{b}. In Fig.~\ref{fig:ObD}\textbf{a}, the Goldstone mode has linear dispersion around the $K_1$-point in the second BZ, corresponding to AFM order. As seen from Fig.~\ref{fig:ObD}\textbf{b}, the mode terminates at the two-Majorana continuum boundary at higher frequency, while another magnon branch tends to the $\Gamma$-point with vanishing intensity. We also extract the spin wave velocity $u\approx 1.15 Ja$ from Fig.~\ref{fig:ObD}\textbf{a}. This is close to the LSWT result $u = (3\sqrt{2}/4) Ja \approx 1.06 Ja$, which is derived in Supplementary Note~\ref*{SM-sec:LSWT}. At high frequencies, the magnon modes are broadened by decaying into the Majorana continua contrary to LSWT.
However, we note that the Majorana MF works less well for the FM Heisenberg model. We present and discuss the results in Supplementary Note~\ref*{SM-sec:SM:Heisenberg}.

We now study the response as a function of the anisotropic parameter $K$, keeping $J>0$ fixed. The parameters and the INS intensities are shown in Fig.~\ref{fig:ObD}\textbf{c}-\textbf{d}. In both cases considered, the system has AFM order. 
As $K$ becomes non-zero the magnon modes acquire a gap $\Delta \sim K$, and the magnetic moments $\mathbf{m}$ are along one of the cubic axes $x,y$ or $z$. The conclusion is true for all phases of the Kitaev-Heisenberg model; we fix $\mathbf{m}\parallel(001)$. 
Let us compare our results with the standard SWT. This is particularly interesting as the Kitaev-Heisenberg model belongs to a class of spin models where ObD effects are present. It means that even though the Kitaev interactions break spin rotational symmetry, the classical ground state still has the full rotational symmetry of the Heisenberg model which is accidental. 
The ObD fluctuation corrections using non-linear SWT (NLSWT) select the magnetic moment direction along the cubic axes as in our approach~\cite{chaloupka2010kitaev,chaloupka2013zigzag}. Interestingly, the correct magnetic order is in our parton theory already determined on the bare MF level, contrary to SWT where magnon interactions need to be considered. This is because the accidental degeneracy is an artifact of the classical spin Ansatz, which the Majorana fermion representation does not use. This difference will become more prominent when we discuss the effect of $\Gamma$ coupling in Sec.~\ref{sec:KJGamma}, where the classical ground state can be a bad approximation in certain parameter regimes.

ObD effects are also visible in the excitation spectra where accidental soft modes from the spurious symmetry enhancement are lifted by quantum fluctuations~\cite{villain1980order,shender1982afm,rau2018pseudo,khatua2023pseudo,rao2025order}. From the excitations spectra we we also extract the magnon gap $\Delta$ for the parameters shown in Fig.~\ref{fig:ObD}, and compare the results with those obtained by ObD SWT fpr $S=1/2$~\cite{rau2018pseudo}: 
\begin{equation}\label{eq:ObD-gap}
\begin{split}
    &K=-0.5J, \ \Delta = 0.2 J \ (\text{MF}), \ \Delta = 0.31 J \ (\text{ObD});\\
    &K=2J, \ \Delta = 1.27 J \ (\text{MF}), \ \Delta = 1.17 J \ (\text{ObD}).
\end{split}
\end{equation}
The good agreement suggests that the quantum fluctuations are already incorporated on the Majorana MF level. In Supplementary Note~\ref*{SM-sec:SM:Heisenberg}, we also find $\Delta$ for $J<0$ and non-zero $K$. There we find that the Majorana MF  gives much larger $\Delta$ than ObD and numerical DMRG results~\cite{gohlke2017dynamics}. This can be attributed to the fact that the FM order is close to being `classical': quantum fluctuations and the associated ObD-induced gap are small and overestimated in the parton treatment.


Similarly, Majorana MFT automatically gives the fluctuation reduction of the magnetic moment from its nominal value $1/2$. At $K=0$, the MF moment $|\mathbf{m}|\approx 0.4$, a reduction by approximately $20\%$. In comparison, linear SWT gives a larger reduction $\Delta m\approx 0.24$; see Supplementary Note~\ref*{SM-sec:LSWT}. This indicates that fluctuations are weaker in the Majorana basis. We note in passing that this value is the same for both $J\gtrless 0$, since they differ by a gauge transformation in the Majorana basis. For $J<0$, this disagrees with the maximally polarized real ground state which is also the classical ground state. For $J>0$ both the magnetic moment and the INS peak intensity decrease with increasing $K$, signifying greater frustration.

\begin{figure}
    \centering
    \includegraphics[width=\linewidth]{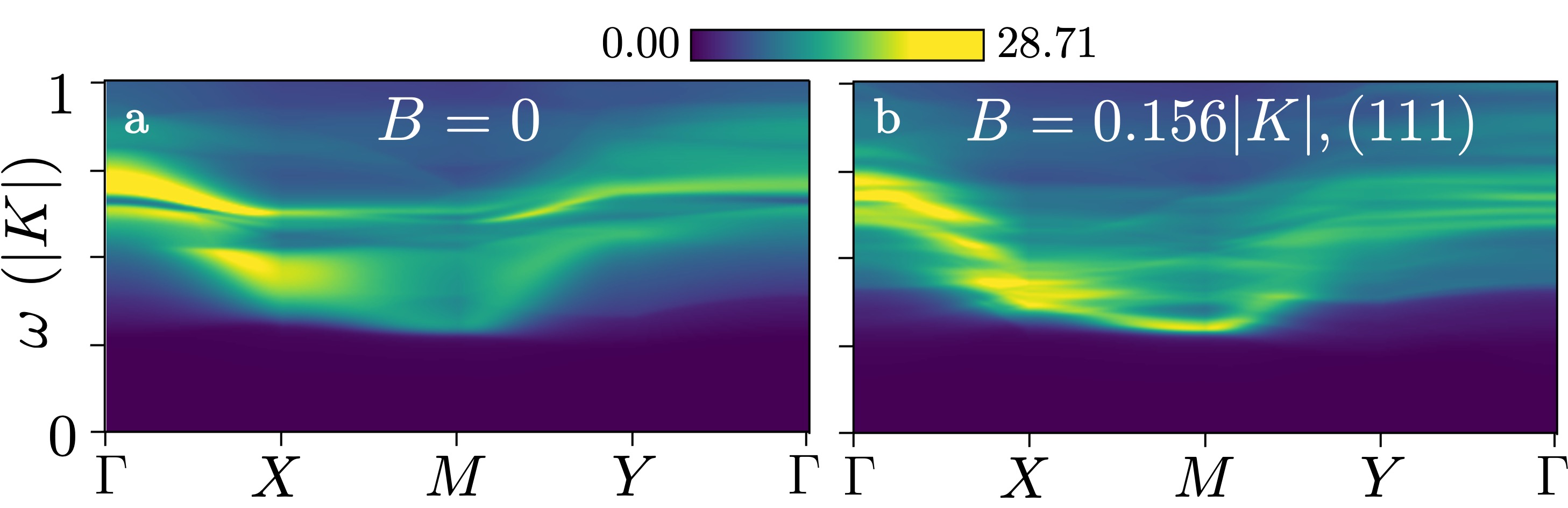}
    \caption{\textbf{INS intensity for $K<0$ and $J = -0.11 |K|$ at external magnetic field.} \textbf{a} $B=0$ and \textbf{b} $B= 0.157 |K|, \ \mathbf{B}\parallel (111)$. It was shown in Ref.~\cite{rao2025dynamical} that for the field value at \textbf{b} the KQSL is stabilized.  }
    \label{fig:proximateQSL}
\end{figure}

We also investigate the stripy phase at $K<0, J = 0.11 |K|$ which is shown to be a `field-induced KQSL'~\cite{rao2025dynamical}, i.e. the condensed magnon modes at $M$-, $M_1$-points at zero field are lifted at intermediate magnetic field values, stabilizing the KQSL. In Fig.~\ref{fig:proximateQSL}, it is shown that the INS intensities in the stripy phase are broad and remain qualitatively the same as the magnetic field is switched on. This signifies that the finite field transition can be first-order, which is expected since the system symmetry is broken explicitly by the magnetic field.

\begin{figure*}
    \centering
    \includegraphics[width=\linewidth]{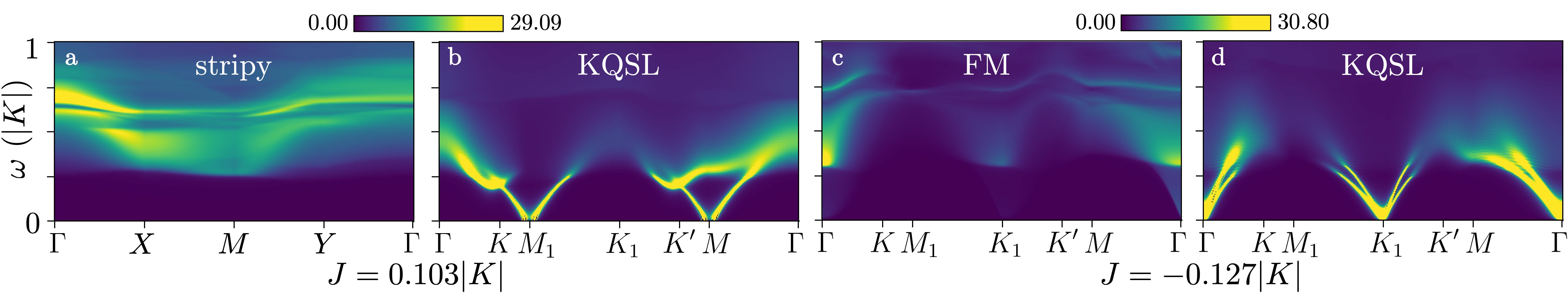}
    \caption{\textbf{INS intensity near the transition points for the Kitaev-Heisenberg model with FM $K$.} Due to proximity to the critical $J_{\text{cr}}$, both phases admit MF solutions. \textbf{a} in the stripy phase and \textbf{b} the QSL phase at $J = 0.103|K|$. \textbf{c} in the FM phase and \textbf{d} the QSL phase at $J = -0.127|K|$. The KQSL phase intensity is normalized to be the same as that in the corresponding ordered phase.  }
    \label{fig:transition}
\end{figure*}

In Ref.~\cite{rao2025dynamical} the critical $J$ values at which the magnon modes condense within the KQSL phase are reported. Here we focus on the transition points for FM $K$: $J_{\text{cr}} \approx -0.12|K|$ and $J_{\text{cr}} \approx 0.1|K|$, which separates the KQSL phase from the ferromagnetic and stripy phases respectively. We plot the corresponding INS intensities in the ordered phases in Fig.~\ref{fig:transition}
Somewhat counter-intuitively, near the transition points the magnon modes are sharper in the KQSL phase than in the magnetically ordered phases. As is seen in Fig.~\ref{fig:transition}, after the on-set of magnetic order the magnons are immediately pushed up into the continuum and become broadened. However, the sudden change is partly a numerical artefact, i.e. the MF solution of the ordered phases does not converge near the critical point. As a result, when the solution has converged at increased $|J|$, there is a first-order like jump to a sizable $\mathbf{m}$.

\subsection{The $KJ\Gamma$-model}\label{sec:KJGamma}
Next, we consider the effect of $\Gamma>0$. Generally, the direction of MF magnetic moments $\mathbf{m}$ are in good agreement with numerical studies for sufficiently large $\Gamma$. In phase $4$ in Fig.~\ref{fig:phases}b, the MF AFM $\mathbf{m}$ lies along the $(111)$-surface which agrees with the classical solution~\cite{rau2014generic}. Phase $3$ has the zigzag moments slightly canted away from the cubic axes, the two inequivalent moments differing by one of the mirror reflections $\sigma$. The $120^\circ$ order for set $6$ has three pairs of AFM moments on adjacent sublattices, differing by $C_6$ rotations. Here however, each pair of AFM moments lies close to one of the $x,y,z$-axes instead of inside the $(111)$-plane predicted by the classical solution. Increasing $\Gamma$ tilts these moments in-plane perpendicular to the bonds. The discrepancy is possibly due to the proximity to the zigzag phases.

Significantly, we find that for small $\Gamma$ the Majorana MF order is not close to the classical ground state. We argue that this should also be the case for the real ground state. Consider for example adding an infinitesimal $\Gamma>0$ coupling inside the FM phase. The classical spin directions are inside the $(111)$-plane and degenerate~\cite{rau2014generic}. However, this is clearly not the correct ground state, given the magnetic moment at $\Gamma = 0$ lies along one of the cubic axes~\cite{chaloupka2010kitaev}, and cannot change discontinuously upon infinitesimal $\Gamma$. The Majorana MF gives much more physical results in comparison: here as $\Gamma$ increases, the magnetic moments are tilted progressively in-plane towards the $(11\bar{2})$-axis, until the system transitions into the zigzag phase. Similarly, classical spins align along the $(111)$-direction for infinitesimal $\Gamma>0$ inside the AFM phase. The Majorana MF shows that within the AFM phase a small $\Gamma>0$ merely tilts the magnetic moment towards the $(111)$-axis, before a first order transition occurs at intermediate $\Gamma$  (around $0.07 K$ for $J=0.2 K$) above which $\mathbf{m}\parallel (111)$. Therefore, here the Majorana MFT can be more convenient, as it captures that the real ground state is far from the classical one.

As for the INS intensities, the $\Gamma$ term generally tends to broaden the magnon mode. As seen from Fig.~\ref{fig:phases}\textbf{b}, finite $\Gamma$ broadens the two-Majorana continuum to lower energies which covers the AFM mode at $K_1$. Within Majorana MF, this is because the $\Gamma$ interaction increases $b$-Majorana level repulsions in the MF spectrum, which generates the broadening effect in the two-Majorana continuum. A similar broadening is observed using NLSWT and DMRG~\cite{winter2017breakdown}, where frustration induced by the anisotropic $\Gamma$ term results in non-conservation of magnon numbers, which gives rise to a broad multi-magnon continuum and magnon decay. There of course, the low energy magnon modes are always sharp. In the presence of a spinon continuum magnon modes can be overdamped over the whole frequency range because there are no kinematic constraints directly linking the single magnon and multi-spinon excitations.

\subsection{Kitaev candidate materials}

We also investigate the `proximate spin liquid' which is argued to be realized in the Kitaev candidate material $\alpha$-RuCl$_3$. 

\begin{figure*}
    \centering
    \includegraphics[width=0.75\linewidth]{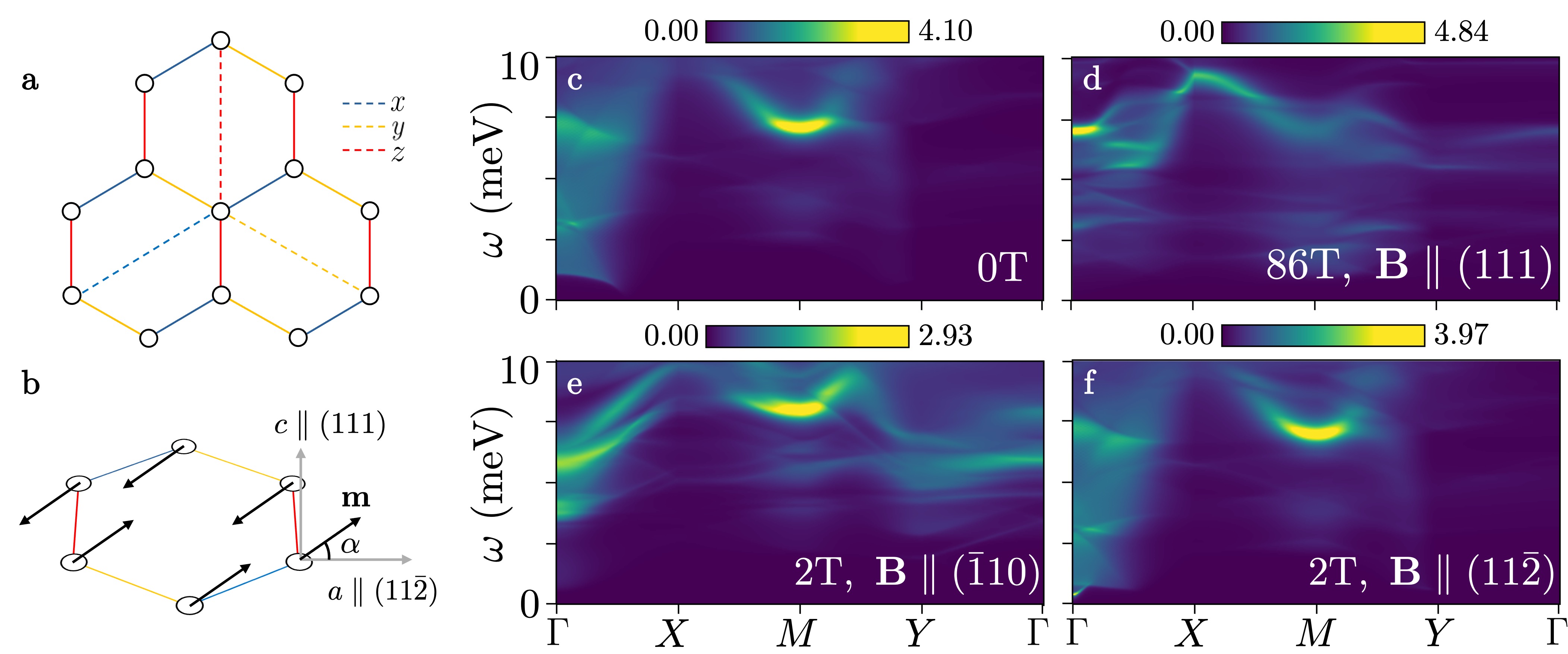}
    \caption{\textbf{The $\mathbf{KJ\Gamma \Gamma' J_3}$-model.} \textbf{a} The third N.N. $x,y,z$-bonds in dashed lines. \textbf{b} Magnetic moments of the zigzag order lying in the $ac$-plane with tilt angle $\alpha$. \textbf{c} INS intensity at zero field. \textbf{d}-\textbf{f} INS intensities for field directions along $a$-, $b$-, $c$-axes. The discontinuity near $\Gamma$ is due to that the INS intensity~\eqref{eq:INS-intensity} does not tend to a constant limit at $\mathbf{k}\rightarrow 0$.}
    \label{fig:candidatematerial}
\end{figure*}

It is commonly believed that $\alpha$-RuCl$_3$ is described by the $KJ\Gamma \Gamma' J_3$-model~\cite{moller2025saga}:
\begin{equation}
    H = H_{KJ\Gamma} + \Gamma'\sum_{\substack{i,j \in \alpha \\ \beta \ne \alpha }} \left[S^\alpha_iS^\beta_j+ S^\beta_i S^\alpha_j\right]+ J_3 \sum_{\text{3rd N.N.}} \mathbf{S}_i.\mathbf{S}_j.
\end{equation}
Here $J_3$ is the third N.N. Heisenberg coupling shown in Fig.~\ref{fig:candidatematerial}\textbf{a}. For microscopic parameters we use the set from Ref.~\cite{moller2025saga} (in units of meV):
\begin{equation}\label{eq:parameters}
    K = - 7.567, J = -4.75, \Gamma= 4.276, \ \Gamma'=2.362, \ J_3 = 3.4. 
\end{equation}

Without magnetic field, the MF ground state has zigzag order where $\mathbf{m}$ lies within the $ac$-plane but is tilted away from the honeycomb plane by angle $\alpha$, agreeing with experiments; see Fig.~\ref{fig:candidatematerial}\textbf{b}. For the parameter set \eqref{eq:parameters}$, \alpha \approx 44.7^\circ$ compared with the experimental value $32^\circ$-$35^\circ$~\cite{moller2025saga}. The INS intensity is shown in Fig.~\ref{fig:candidatematerial}\textbf{c}. There is a relatively broad peak near the $M$ point of the zigzag order, and a large continuum response near $\Gamma$. The overall shape of the response is in agreement with the INS data~\cite{banerjee2017neutron} and the broad continua stem from two-partcile Majorana excitations. 

We also consider the effect of an external magnetic field along the $a,b,c$-axes. We use a simple isotropic $g$-factor $g=2$. In Fig.~\ref{fig:candidatematerial}(d)-(f), we plot the INS intensities just below the critical field values for the KQSL MF solution. In particular under $\mathbf{B}\parallel(\bar{1}10)$, the moment $\mathbf{m}$ is strongly canted towards $(\bar{1}10)$ under modest field values. The large anisotropy in critical field values $B_{\text{cr}} \approx 86~$T along the $c$-axis and around $2$~T along in-plane $a$-and $b$-axes, agrees qualitatively with experiments ($|\mathbf{B}_{\text{cr}}|\approx 8~$T in plane and $70~$T out-of-plane~\cite{trebst2022kitaev}). A more quantitative determination of $\mathbf{B}_{\text{cr}}$ to compare with experiments however, would require the precise form of the anisotropic $g$-tensor of the material and correction to the mean field values from fluctuations.

\section{Discussion}

In this work, we employed a self-consistent Majorana MFT and RPA theory to study the magnetization and INS intensity in the magnetically ordered phases of the $KJ\Gamma$-model on the honeycomb lattice. Motivated by the concept of the 'proximate QSL' in $\alpha-$RuCl$_3$ our approach not only captures the low energy magnon like excitations but also the higher energy fractionalized excitations of a nearby QSL phase, which can overdamp magnon modes. We find overall good qualitative agreement between the phases obtained by our MF theory and other numerical methods. Remarkably, at $\Gamma=0$ and $K\ne 0$ we find that Majorana MFT automatically gives the correct ground state spin configurations and a finite spin gap, which normally requires ObD fluctuation corrections beyond linear SWT. We also confirmed that in the AFM Heisenberg limit, the spin isotropy and the linear Goldstone modes are reproduced correctly within our approach with quantitative agreement of the spin wave velocity. For the $KJ\Gamma$-model, we found gapped magnon modes at low energy in the ordered phase with minima near the corresponding high-symmetry points of the BZ. The magnons are generally broadened near the phase boundary with the KQSL and only become sharp upon increased Heisenberg coupling. The $\Gamma$ coupling however, tends to extend the Majorana continuum to lower energy, thus broadening the magnon modes as a result. In the NLSWT formalism, $\Gamma$ has a similar effect on the single-magnon branch by increasing frustration and inducing magnon non-conserving processes. This facilitates magnon decay into multi-magnon continuum and the single-magnon modes can become broadened at higher energies. In terms of Majorana fermions, the broadening is due to magnons decaying into the multi-Majorana continuum. One crucial difference between the two methods is that the lowest energy magnon modes are always sharp in NLSW due to kinematic constraints on magnon decay. However, in the Majorana RPA the magnon modes can be broad throughout the BZ because the kinematics of the continuum which damps them is not linked to the magnon dispersion itself. 

In this context, we add a word of caution. While the parton RPA method captures part of the gauge field fluctuations of the QSL it does not correctly describes the confinement of the two Majorana sector. This is neither a problem at low energies, where the RPA recovers the collective magnon modes as Majorana bound states, nor at high energies, where the short time dynamics is unaffected by coonfinement. However, the bottom of the two-Majorana continuum is expected to be altered and form confinement-induced sharp modes similar to the established Ising chain materials~\cite{rutkevich2010weak}. Extending the theory to capture this crossover regime is a formidable challenge for future research. 

We also investigated the $KJ\Gamma\Gamma' J_3$-model which is argued to describe the Kitaev candidate material $\alpha$-RuCl$_3$. The spin coupling parameters proposed in Ref.~\cite{moller2025saga} gives a MF zigzag phase that lies in the plane of the $a$-, $c$-axes with $\mathbf{m}$ tilted from the $a$-axis by $\alpha \approx 44.7^\circ$. This agrees qualitatively with the experimental value of $32^\circ-35^\circ$. We also reproduce qualitatively the large anisotropy in the stability of the proximate QSL with respect to external field directions. The INS intensities are concentrated near $M$ and $\Gamma$ in the BZ, but there are no sharp modes, again similar to experimental observations~\cite{banerjee2017neutron,sarkis2026intermediate}.

Overall, our method predicts quantitatively key features of the INS signature, i.e. sharp modes and the broad continua, in frustrated magnetic systems. Compared with other approaches, our framework is general and self-consistent without ad hoc approximations and free parameters. Requiring only the spin couplings as input, both the continua and the sharp modes are generated by Majorana fermions, which resembles stable and unstable bound stats. The considerable simplicity and generality is a methodological advantage of our approach and will help to interpret both experimental data and numerical results. 

One possible extension is to consider other types of responses. In this paper and in Ref.~\cite{rao2025dynamical} we have considered the dynamical response of spin operators related to INS experiments, which captures the spin-$1/2$ quasiparticle spectrum. The continuum appears indirectly as a result of quasiparticle decay. However to directly probe the continuum, other types of dynamical response functions can be more convenient. For example, Raman and resonant inelastic X-ray scattering (RIXS) experiments measure the dynamical response of two-spin operators and capture the continuum directly~\cite{knolle2014raman,nasu2016fermionic,halasz2016resonant}. Thus investigating such response functions using the Majorana RPA is an interesting direction for the future. Similarly, our method can be readily applied to other ordered phases proximate to a QSL phase, for example the undoped cuprates~\cite{ho2001nature}, the Kagome lattice QSL~\cite{iqbal2011projected}, phases proximate to the U(1) Dirac QSL on the triangular lattice~\cite{willsher2025dynamics}, and the square lattice $J_1$-$J_2$ model which is argued to host a $\mathbb{Z}_2$ QSL~\cite{ferrari2020gapless,shackleton2021deconfined}. Another exciting option is to consider other phases with fractionalized excitations well described by parton theories, such as the fractional Quantum Hall effect~\cite{stormer1999fqhe} and the fractional Chern insulators~\cite{regnault2011fractional}. The simple and intuitive nature of RPA could potentially provide a better understanding of these systems and allow for a crisp understanding of experimental observables.


\section{Methods}

\subsection{Majorana mean field theory}

The mapping from spin-$1/2$ operators to Majorana fermions~\cite{kitaev2006anyons}:
\begin{equation}\label{eq:spin-mapping}
    S^\alpha_i \rightarrow i c_i b^\alpha_i, \ c_i^2 = (b^\alpha_i)^2=\frac{1}{2},
\end{equation}
with the on-site quartic constraint:
\begin{equation}\label{eq:quartic-constraint}
     D_i  = c_ib^x_ib^y_ib^z_i = \frac{1}{4}.
\end{equation}

The Majorana band structure is obtained using Majorana mean field theory (MFT) by decoupling in the following spin-liquid channel:
\begin{equation}
    \eta_\alpha = i \langle c_i c_j\rangle_\alpha,\ Q^{\beta\gamma}_\alpha = i \langle b^\beta_i b^\gamma_j\rangle_\alpha, \ m^\alpha_i = i\langle c_i b^\alpha_i\rangle, 
\end{equation}
where $i,j\in \alpha$ are connected by the N.N. $\alpha$-bond. The quartic constraint \eqref{eq:quartic-constraint} can be written equivalently as three quadratic constraints~\cite{you2012doping}: 
\begin{equation}
    F^\alpha_i = i\left(c_ib^\alpha_i + \frac{1}{2} \varepsilon^{\alpha\beta\gamma} b^\beta_i b^\gamma_i\right).
\end{equation}
They are enforced on average $\sum_i \langle F^\alpha_i\rangle =0$ by adding the terms $\lambda_\alpha F^\alpha_i$ to the MF Hamiltonian, where $\lambda_\alpha$ are the Lagrange multipliers regarded as additional MF parameters.

For the $KJ\Gamma\Gamma'J_3$-model, there appears additional MF parameters for sites $i,j$ connected by third N.N. bonds:
\begin{equation}
    \widetilde{\eta}_{\alpha'}= i \langle c_i c_j\rangle_{\alpha'},\ \widetilde{Q}^{\beta\beta}_{\alpha'} = i \langle b^\beta_i b^\beta_j\rangle_{\alpha'},
\end{equation}
where $\alpha' = x,y,z$ as shown in Fig.~\ref{fig:candidatematerial}a. 

The MF Majorana Hamiltonian in the magnetically ordered phases are obtained by first finding the MF solution in the KQSL regime, then solving for $m_i^\alpha$ keeping all other MF parameters fixed. 

More precisely, \textit{within the QSL phase} at zero magnetic field, we look for the $m^\alpha_i=0$ solution with the two-sublattice unit cell, which always exists. The magnetic field induces non-zero $\mathbf{m}$. For field directions $\mathbf{B}\parallel (111)$ and $\mathbf{B}\parallel (\bar{1}10)$, the system has six-fold rotation symmetry $\mathrm{C}_6$ and a reflection plane $\sigma$ across the $z$-bond respectively. This allows the number of independent MF parameters to be reduced in the same way as described in Ref.~\cite{you2012doping}. Note that the third N.N. bonds transform in the same way as N.N. bonds under the symmetry operations so the same consideration applies to $\widetilde{\eta}_{\alpha'}, \widetilde{Q}^{\beta\beta}_{\alpha'}$. 

For the \textit{magnetically ordered phases}, the spontaneous magnetization is found by solving for $m^\alpha_i$ while keeping all other MF parameters fixed. For FM and AFM, the unit cell is unaltered. For zigzag and stripy phases, we use the four-sublattice unit cell, while for the $120^\circ$ the unit cell has six sublattices; see Fig.~\ref{fig:unitcell}.

For all MF simulations, we use the momentum discretization:
\begin{equation}\label{eq:momentum-discrete}
    \mathbf{p} = (n_1\mathbf{b}_1 + n_2\mathbf{b}_2 )/L,
\end{equation}
where $0\leq n_i <L, L=60$ and $\mathbf{b}_i$ are the corresponding reciprocal lattice vectors in a given unit cell.

\subsection{Spin susceptibility}\label{sec:spin-susceptibility}

The spin susceptibility is computed using RPA. First we compute the one-loop susceptibility tensor in the Matsubara formalism at zero temperature $T=0$:
\begin{equation}\label{eq:one-loop-tensor}
    [\chi_0 (\omega,\mathbf{k})]_{i\alpha,j\beta} = \lim_{ik_0 \rightarrow \omega+i\delta}\int  \big\langle \text{T}_\tau \{S^\alpha_i(\tau,\mathbf{k})S^\beta_j(0,-\mathbf{k})\} \big\rangle e^{ik_0 \tau} \diff \tau,
\end{equation}
where $i,j$ are sublattice indices and $S^\alpha_i(\tau,\mathbf{k})$ is the Fourier transform of the spin operator at momentum $\mathbf{k}$ in the Majorana basis~\eqref{eq:spin-mapping}. The averaging is taken over the non-interacting MF background. $\delta$ is a positive infinitesimal in the analytical continuation $ik_0 \rightarrow \omega +i\delta$. Eq.~\eqref{eq:one-loop-tensor} can be computed by diagonalizing the MF Majorana Hamiltonian in momentum space, taking the eigenstates and energy levels as input. This was done in Ref.~\cite{rao2025dynamical} for the two-sublattice case without magnetic order. To generalize to larger unit cells, the indices $i,j$ include more sublattices and the MF Hamiltonian contains a non-zero $\mathbf{m}$. In choosing the unit-cells as shown in Fig.~\ref{fig:unitcell}, the BZ backfolding due to magnetic order is automatically taken into account.

The four-Majorana interactions are given by substituting the mapping \eqref{eq:spin-mapping} into the given spin Hamiltonian, which acquires the following form in momentum space:
\begin{equation}\label{eq:interaction-vertex}
    V  =  \sum_{ \mathbf{p},\mathbf{p'},\mathbf{k}} U_{i\alpha,j\beta}(\mathbf{k}) i c_i(\mathbf{k}- \mathbf{p}) b_i^\alpha(\mathbf{p})  i c_j(-\mathbf{k}- \mathbf{p}') b_j^\beta(\mathbf{p}').
\end{equation}
The RPA sums over all one-loop diagrams with the interaction vertex~\eqref{eq:interaction-vertex}. The resulting susceptibility tensor is given by the `matrix equation':
\begin{equation}\label{eq:RPA-tensor}
    \chi (\omega,\mathbf{k})  = \chi_0(\omega,\mathbf{k}) \left[1+\hat{U}(\mathbf{k})\chi_0(\omega,\mathbf{k})\right]^{-1},
\end{equation}
where $ [\hat{U}(\mathbf{k})]_{i\alpha,j\beta}  = U_{i\alpha,j\beta} (-\mathbf{k}) + U_{j\beta,i\alpha}(\mathbf{k})$ is the symmetrized interaction vertex. The spin susceptibility is:
\begin{equation}\label{eq:RPA-susceptibility}
    \chi_{\alpha\beta}(\omega,\mathbf{k}) = \sum_{i,j}[\chi_{\alpha\beta}(\omega,\mathbf{k})]_{i\alpha,j\beta}e^{-i\mathbf{k}.(\mathbf{r}_i-\mathbf{r}_j)},
\end{equation}
where $\mathbf{r}_i-\mathbf{r}_j$ are vectors connecting different sublattices in a single unit cell. For example, for the two-sublattice unit cell: $\mathbf{r}_B -\mathbf{r}_A = -(\mathbf{a}_1+\mathbf{a}_2)/3$.

Finally, from the spin susceptibility we compute the INS intensity:
\begin{equation}\label{eq:INS-intensity}
    \mathcal{I}(\omega,\mathbf{k}) =\left(\delta_{\alpha\beta} -\frac{k_\alpha k_\beta}{k^2}\right) i\left[\chi_{\beta\alpha}^*(\omega,\mathbf{k})-\chi_{\alpha\beta}(\omega,\mathbf{k})\right],
\end{equation}
where $k_\alpha, k_\beta$ are three-dimensional momenta in the global laboratory frame. The factor $i[\chi_{\beta\alpha}^*(\omega,\mathbf{k})-\chi_{\alpha\beta}(\omega,\mathbf{k})]$ represents the dynamical structure factor due to the fluctuation-dissipation theorem. For numerical simulations, we use $L = 180$ in \eqref{eq:momentum-discrete} and $\omega = n \omega_{\text{max}}/L, \ 0 \leq n<L$ where $\omega_{\text{max}}$ is the upper frequency limit shown in each plot. The broadening $\delta = \omega_{\text{max}}/L$. The logarithmic scale for INS plots is given by $\log (c+\mathcal{I}/40)$ with $c=1$ unless stated otherwise.

\section*{Data availability}

The numerical code for computing the Majorana MF spectrum and the spin susceptibility is available at \cite{code}.

\section*{Acknowledgements} 

We would like to thank K. Dixit, C. Sarkis, A. Banerjee, C. Balz, S. Nagler and A. Tennant for a related experimental collaboration. JK acknowledges support from the Deutsche Forschungsgemeinschaft (DFG, German Research Foundation) under Germany’s Excellence Strategy (EXC–2111–390814868 and ct.qmat EXC-2147-390858490),
and DFG Grants No. KN1254/1-2, KN1254/2-1 TRR 360 - 492547816 and SFB 1143 (project-id 247310070), as well as the Munich Quantum Valley, which is supported by the Bavarian state government with funds from the Hightech Agenda Bayern Plus.

\nobalance 


%

\end{document}


\title{Supplemental Material for 'Proximate Quantum spin liquids and Majorana continua in magnetically ordered Kitaev magnets'}

\author{Peng Rao}
\affiliation{Physics Department, Technical University of Munich, TUM School of Natural Sciences, 85748 Garching, Germany}
\author{Roderich Moessner}
\affiliation{Max Planck Institute for the Physics of Complex Systems, 01187 Dresden, Germany}
\author{Johannes Knolle}
\affiliation{Physics Department, Technical University of Munich, TUM School of Natural Sciences, 85748 Garching, Germany}
\affiliation{Munich Center for Quantum Science and Technology (MCQST), Schellingstr. 4, 80799 München, Germany}
\affiliation{Blackett Laboratory, Imperial College London, London SW7 2AZ, United Kingdom}

\date{\today}
\begin{abstract}
\tableofcontents
\end{abstract}

\maketitle

\section{Linear spin wave theory in the AFM Heisenberg limit}\label{sec:LSWT}

Applying the Holstein-Primakoff expansion to the AFM Heisenberg Hamiltonian, the quadratic magnon Hamiltonian is:
\begin{equation}\label{eq:LSWT-Hamiltonian}
    H_{\text{mag}} =  JSz \sum_{\mathbf{k}} \Psi_\mathbf{k}^\dagger  \begin{pmatrix} 1 & \gamma_\mathbf{k} \\ \gamma_{\mathbf{k}}^* & 1    \end{pmatrix} \Psi_\mathbf{k}, \ \gamma_\mathbf{k} =\frac{1}{z}\sum_{\alpha } e^{i\mathbf{k}.\mathbf{e}_\alpha}, \ \Psi_\mathbf{k} = \begin{pmatrix}  a_\mathbf{k} \\ b_{-\mathbf{k}}^\dagger  \end{pmatrix}, \ 
\end{equation}
where $a_\mathbf{k}, \ b_{\mathbf{k}}$ are magnon operators on sublattices $A$ and $B$, and $z=3$ is the coordination number. In $\gamma_\mathbf{k}$, the summation is over N.N. bond vectors $\mathbf{e}_\alpha$.

Diagonalizing Eq.~\eqref{eq:LSWT-Hamiltonian} using the standard Bogoliubov transformation results in the spectrum:
\begin{equation}
    E_\mathbf{k} = JSz \sqrt{1-|\gamma_\mathbf{k}|^2}.
\end{equation}
Expanding at small $\mathbf{k}$ and $S=1/2, z=3$, we obtain:
\begin{equation}
  \gamma_\mathbf{k} \approx 1 - \frac{(ka)^2}{4}, \   E_\mathbf{k} \approx \frac{3\sqrt{2}}{4} J ka,
\end{equation}
where $a$ is the lattice constant

The fluctuation correction to the on-site magnetic moment is:
\begin{equation}
\Delta m=\frac{1}{2}\sum_\mathbf{k} \left[(1-|\gamma_\mathbf{k}|^2)^{-1/2} - 1\right]\approx 0.24.
\end{equation}

\section{The ferromagnetic Heisenberg model}\label{sec:SM:Heisenberg}

The INS intensities for the Kitaev-Heisenberg model with FM couplings $K<0, J<0$ are presented in Supplementary Figure~\ref{fig:SM:ObD-FM}. It can be seen at once in Supplementary Figure~\ref{fig:SM:ObD-FM}\textbf{a} that the magnon dispersion is linear in the Heisenberg limit. Furthermore, the dispersion is the same as the AFM Heisenberg model in Fig.~\ref*{M:fig:ObD}\textbf{a}, but the sharp mode near $\Gamma$ does not vanish. This contradicts the $\omega\propto k^2$ dispersion of the Heisenberg ferromagnet, which is known to be a general consequence of the conversation of magnetization as the symmetry-breaking order parameter~\cite{halperin1969hydrodynamic}. We attribute the discrepancy to that the MF Majorana Hamiltonian breaks spin conservation and the consideration in Ref.~\cite{halperin1969hydrodynamic} does not apply. In fact, below we shall prove that the magnon dispersions obtained from Majorana RPA are the same for both FM and AFM Heisenberg models.

In the Majorana basis, the FM Heisenberg model differs from the AFM model by a gauge transformation, e.g. on $c$-Majorana fermions on the $B$ sublattices:
\begin{equation}\label{eq:SM:gauge-transform}
    c_B \rightarrow - c_B, \ \mathbf{S}_B \rightarrow -\mathbf{S}_B.
\end{equation}
As a result, the non-interacting spin susceptibility tensor $ [\chi_0(\omega,\mathbf{k})]_{i\alpha,j\beta}$~(\ref*{M:eq:one-loop-tensor}) for given spin indices $\alpha,\beta$ is transformed in the sublattice indices $i,j=A,B$ as follows:
\begin{equation}
    \chi_0 = \begin{bmatrix} (\chi_0)_{AA} &  (\chi_0)_{AB} \\ (\chi_0)_{BA} & (\chi_0)_{BB}   \end{bmatrix} \rightarrow  \begin{bmatrix} (\chi_0)_{AA} &  -(\chi_0)_{AB} \\ -(\chi_0)_{BA} & (\chi_0)_{BB}   \end{bmatrix}
\end{equation}
where we neglect $\alpha, \beta$.

Let us now compute analytically the RPA susceptibility for the AFM Heisenberg model. The interaction vertex Eq.~(\ref*{M:eq:interaction-vertex}) has the form:
\begin{equation}\label{eq:SM:interaction-vertex}
    U_{i\alpha,j\beta}(\mathbf{k}) =\delta_{\alpha\beta} \begin{pmatrix} 0 & J  \gamma_\mathbf{k}\\ J \gamma_\mathbf{k}^*& 0    \end{pmatrix}_{ij}; \ \gamma_\mathbf{k}=\sum_\alpha e^{i\mathbf{k}.\mathbf{e}_\alpha}.
\end{equation}
Thus the RPA equation~(\ref*{M:eq:RPA-tensor}) factorizes into $2\times 2$ matrix equations in sublattice indices only. Thus we shall neglect spin indices below and regard all relevant tensors as $2\times 2$ matrices. Let us first consider the quasi-particle pole given by the zeros of the RPA kernel:
\begin{equation}
    \det[1+\hat{U}(\mathbf{k}) \chi_0] = \det \begin{pmatrix} 1 + J \gamma_\mathbf{k}(\chi_0)_{BA} & J  \gamma_\mathbf{k}  (\chi_0)_{BB}\\ J \gamma_\mathbf{k}^* (\chi_0)_{AA}&1+ J \gamma_\mathbf{k}^* (\chi_0)_{AB}   \end{pmatrix}.
\end{equation}
Under the gauge transformation \eqref{eq:SM:gauge-transform} and $J\rightarrow -J $ in \eqref{eq:SM:interaction-vertex}, the off-diagonal components change sign while the diagonal ones are invariant. Therefore, the determinant is also invariant and both FM and AFM Heisenberg models have the same magnon dispersions. 

The RPA susceptibility and the INS intensity are of course not invariant (for simplicity we neglect the momentum arguments):
\begin{equation}
    \chi = \chi_0(1+\hat{U}\chi_0)^{-1}=\frac{1}{  \det(1+\hat{U} \chi_0) } \begin{pmatrix} (\chi_0)_{AA} & (\chi_0)_{AB} + J\gamma_\mathbf{k}\det \chi_0\\ (\chi_0)_{BA} + J\gamma_\mathbf{k}^*\det \chi_0& (\chi_0)_{BB}  \end{pmatrix}.
\end{equation}
Under the transformation \eqref{eq:SM:gauge-transform} and $J\rightarrow -J $ in \eqref{eq:SM:interaction-vertex}, $\chi$ transforms in the same way as $\chi_0$: only the off-diagonal components change sign ($\det \chi_0$ is clearly invariant). As can be seen from Eqs.~(\ref*{M:eq:RPA-susceptibility}) and (\ref*{M:eq:INS-intensity}), the resulting INS intensity is also not invariant. Near $\mathbf{k}=0$ for the AFM model, the off-diagonal components add destructively to the diagonal ones, and the magnon mode intensity vanishes. For the FM model, they add constructively near $\mathbf{k}=0$ resulting in the observed sharp mode with the same dispersion as the AFM case. We note that in Fig.~\ref*{M:fig:ObD}\textbf{b} the $\mathbf{k}=0$ mode is entirely absent from $\Gamma$ to $K$. This is because the corresponding momentum path has $\mathbf{k}.\mathbf{e}_z= 0$ in the sublattice factor in Eq.~(\ref*{M:eq:RPA-tensor}), so the cancellation holds even away from $\mathbf{k}=0$.

\begin{figure}
    \centering
    \includegraphics[width=0.8\linewidth]{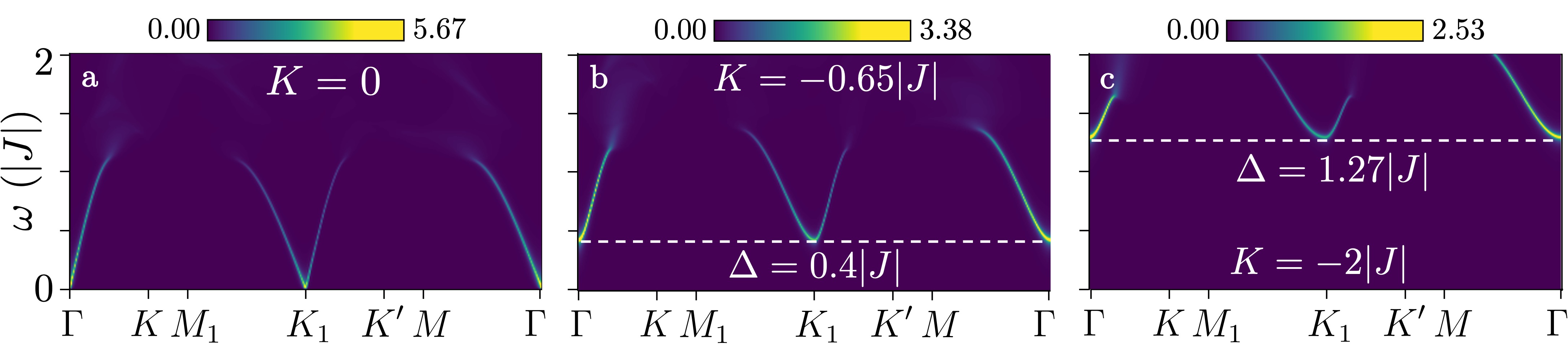}
    \caption{INS intensities for the Kitaev-Heisenberg model inside the FM phase at $J<0$. The intensities are shown in logarithmic scale described in the Method section in the main text. }
    \label{fig:SM:ObD-FM}
\end{figure}

We also extract the magnon gap for the FM models in Supplementary Figure~\ref{fig:SM:ObD-FM}\textbf{b}-\textbf{c} and compare with ObD results~\cite{rau2018pseudo}:
\begin{equation}
\begin{split}
    &K=-0.65 |J|, \ \Delta = 0.4 |J| \ (\text{MF}), \ \Delta = 0.03 |J| \ (\text{ObD});\\
    &K=-2|J|, \ \Delta = 1.27 J \ (\text{MF}), \ \Delta = 0.208 J \ (\text{ObD}).
\end{split}
\end{equation}
Note that for $K = -2|J|$ the MF gap is the same as the AFM model with $K=2J$ in (\ref*{M:eq:ObD-gap}). This is due to that the two models are related by the gauge transformation \eqref{eq:SM:gauge-transform}, and a similar analysis to above shows that the quasi-particle pole is also invariant in this case. The Majorana MF overestimates the magnon gaps compared with both ObD and numerics~\cite{rau2018pseudo,gohlke2017dynamics}.

Finally, we remark that a similar transformation to \eqref{eq:SM:gauge-transform} in the zigzag phase on sublattices $A, C$ (shown in Figure~\ref*{M:fig:unitcell}\textbf{b}) also changes the signs of spin coupling, and converts the system into the stripy phase. This explains the similarity of the INS intensities in Fig.~\ref*{M:fig:proximateQSL} between sets $2$ and $5$.

%